\pdfoutput=1
\documentclass[12pt,a4paper]{article}

\usepackage{ifthen} 
\newboolean{pdflatex}
\setboolean{pdflatex}{false} 

\newboolean{articletitles}
\setboolean{articletitles}{true} 

\newboolean{uprightparticles}
\setboolean{uprightparticles}{false} 

\def\paperauthors{LHCb collaboration} 
\def\paperasciititle{Measurement of the Z-boson mass} 
\def\papertitle{Measurement of the $Z$-boson mass} 
\def\paperkeywords{{High Energy Physics}, {LHCb}} 
\def\papercopyright{\the\year\ CERN for the benefit of the LHCb collaboration} 
\def\paperlicence{CC BY 4.0 licence}
\def\paperlicenceurl{https://creativecommons.org/licenses/by/4.0/}

\newif\ifEnableSectionTOCLinks
\EnableSectionTOCLinksfalse 

\usepackage[top=1in, bottom=1.25in, left=1in, right=1in]{geometry}

\usepackage{microtype}
\usepackage{lineno}  
\usepackage{xspace} 
\usepackage{caption} 

\usepackage{graphicx}  
\usepackage{color}
\usepackage{colortbl}
\graphicspath{{./figs/}} 

\usepackage{amsmath} 
\usepackage{amssymb}
\usepackage{amsfonts}
\usepackage{upgreek} 

\newcommand*\patchAmsMathEnvironmentForLineno[1]{%
\expandafter\let\csname old#1\expandafter\endcsname\csname #1\endcsname
\expandafter\let\csname oldend#1\expandafter\endcsname\csname
end#1\endcsname
 \renewenvironment{#1}%
   {\linenomath\csname old#1\endcsname}%
   {\csname oldend#1\endcsname\endlinenomath}%
}
\newcommand*\patchBothAmsMathEnvironmentsForLineno[1]{%
  \patchAmsMathEnvironmentForLineno{#1}%
  \patchAmsMathEnvironmentForLineno{#1*}%
}
\AtBeginDocument{%
\patchBothAmsMathEnvironmentsForLineno{equation}%
\patchBothAmsMathEnvironmentsForLineno{align}%
\patchBothAmsMathEnvironmentsForLineno{flalign}%
\patchBothAmsMathEnvironmentsForLineno{alignat}%
\patchBothAmsMathEnvironmentsForLineno{gather}%
\patchBothAmsMathEnvironmentsForLineno{multline}%
\patchBothAmsMathEnvironmentsForLineno{eqnarray}%
}

\usepackage[pdftex,
            pdfauthor={\paperauthors},
            pdftitle={\paperasciititle},
            pdfkeywords={\paperkeywords}]{hyperref}
\usepackage{hyperxmp}
\hypersetup{
    pdfcopyright={Copyright (C) \papercopyright},
    pdflicenseurl={\paperlicenceurl}
}

\usepackage[colorinlistoftodos,textsize=scriptsize]{todonotes}

\usepackage[bottom,flushmargin,hang,multiple]{footmisc}

\usepackage[all]{hypcap} 

\usepackage{xspace} 
\usepackage{upgreek}

\def\lhcb   {\mbox{LHCb}\xspace}

\def\lhc    {\mbox{LHC}\xspace}

\def\MagUp {\mbox{\em Mag\kern -0.05em Up}\xspace}

\ifthenelse{\boolean{uprightparticles}}%
{

 \def\Pmu         {\ensuremath{\upmu}\xspace}

 \def\Ptau        {\ensuremath{\uptau}\xspace}

 \def\Ppsi        {\ensuremath{\uppsi}\xspace}

 \def\PDelta      {\ensuremath{\Delta}\xspace}                 
 \def\PXi         {\ensuremath{\Xi}\xspace}                 
 \def\PLambda     {\ensuremath{\Lambda}\xspace}                 
 \def\PSigma      {\ensuremath{\Sigma}\xspace}                 
 \def\POmega      {\ensuremath{\Omega}\xspace}                 
 \def\PUpsilon    {\ensuremath{\Upsilon}\xspace}
 \let\oldPi\Pi
 \def\PPi         {\ensuremath{\oldPi}\xspace}

 \def\PB      {\ensuremath{\mathrm{B}}\xspace}                 
 \def\PD      {\ensuremath{\mathrm{D}}\xspace}                 
 \def\PJ      {\ensuremath{\mathrm{J}}\xspace}                 
 \def\PK      {\ensuremath{\mathrm{K}}\xspace}                 
 \def\PW      {\ensuremath{\mathrm{W}}\xspace}                 
 \def\PZ      {\ensuremath{\mathrm{Z}}\xspace}                 
 \def\Ps      {\ensuremath{\mathrm{s}}\xspace}

 \def\thebaroffset{0.0em}
}
{

 \def\Pmu         {\ensuremath{\mu}\xspace}

 \def\Ptau        {\ensuremath{\tau}\xspace}

 \def\Ppsi        {\ensuremath{\psi}\xspace}                 
                  
 \mathchardef\PDelta="7101
 \mathchardef\PXi="7104
 \mathchardef\PLambda="7103
 \mathchardef\PSigma="7106
 \mathchardef\POmega="710A
 \mathchardef\PUpsilon="7107
 \mathchardef\PPi="7105
 \def\PB      {\ensuremath{B}\xspace}                 
 \def\PD      {\ensuremath{D}\xspace}                 
 \def\PJ      {\ensuremath{J}\xspace}                 
 \def\PK      {\ensuremath{K}\xspace}                 
 \def\PW      {\ensuremath{W}\xspace}                 
 \def\PZ      {\ensuremath{Z}\xspace}                 
 \def\Ps      {\ensuremath{s}\xspace}

 \def\thebaroffset{0.18em}
}
\newcommand{\offsetoverline}[2][\thebaroffset]{\kern #1\overline{\kern -#1 #2}}%

\makeatletter
\ifcase \@ptsize \relax
  \newcommand{\miniscule}{\@setfontsize\miniscule{4}{5}}
\or
  \newcommand{\miniscule}{\@setfontsize\miniscule{5}{6}}
\or
  \newcommand{\miniscule}{\@setfontsize\miniscule{5}{6}}
\fi
\makeatother

\DeclareRobustCommand{\optbar}[1]{\shortstack{{\miniscule (\rule[.5ex]{1.25em}{.18mm})}
  \\ [-.7ex] $#1$}}

\def\mup        {{\ensuremath{\Pmu^+}}\xspace}
\def\mun        {{\ensuremath{\Pmu^-}}\xspace} 

\def\mumu       {{\ensuremath{\Pmu^+\Pmu^-}}\xspace}

\def\tautau     {{\ensuremath{\Ptau^+\Ptau^-}}\xspace}

\def\squark    {{\ensuremath{\Ps}}\xspace}

\def\KorKbar {\kern \thebaroffset\optbar{\kern -\thebaroffset \PK}{}\xspace}

\def\D       {{\ensuremath{\PD}}\xspace}

\def\DorDbar {\kern \thebaroffset\optbar{\kern -\thebaroffset \PD}\xspace}

\def\Dp      {{\ensuremath{\D^+}}\xspace}
\def\Dm      {{\ensuremath{\D^-}}\xspace}

\def\DpDm    {\ensuremath{\Dp {\kern -0.16em \Dm}}\xspace}

\def\B       {{\ensuremath{\PB}}\xspace}

\def\BorBbar {\kern \thebaroffset\optbar{\kern -\thebaroffset \PB}\xspace}

\def\Bd      {{\ensuremath{\B^0}}\xspace}

\def\BdorBdbar {\kern \thebaroffset\optbar{\kern -\thebaroffset \Bd}\xspace}

\def\Bs      {{\ensuremath{\B^0_\squark}}\xspace}

\def\BsorBsbar {\kern \thebaroffset\optbar{\kern -\thebaroffset \Bs}\xspace}

\def\jpsi     {{\ensuremath{{\PJ\mskip -3mu/\mskip -2mu\Ppsi}}}\xspace}

\def\Upsilonres  {{\ensuremath{\PUpsilon}}\xspace}
\def\Y#1S{\ensuremath{\PUpsilon{(#1S)}}\xspace}

\def\LorLbar     {\kern \thebaroffset\optbar{\kern -\thebaroffset \PLambda}\xspace}

\newcommand{\decay}[2]{\mbox{\ensuremath{#1\!\to #2}}\xspace} 

\def\to                 {\ensuremath{\rightarrow}\xspace}

\def\AT#1     {\ensuremath{A_{\mathrm{T}}^{#1}}\xspace}           

\def\C#1      {\ensuremath{\mathcal{C}_{#1}}\xspace}                       
\def\Cp#1     {\ensuremath{\mathcal{C}_{#1}^{'}}\xspace}                    
\def\Ceff#1   {\ensuremath{\mathcal{C}_{#1}^{\mathrm{(eff)}}}\xspace}        
\def\Cpeff#1  {\ensuremath{\mathcal{C}_{#1}^{'\mathrm{(eff)}}}\xspace}       
\def\Ope#1    {\ensuremath{\mathcal{O}_{#1}}\xspace}                       
\def\Opep#1   {\ensuremath{\mathcal{O}_{#1}^{'}}\xspace}                    

\newcommand{\aunit}[1]{\ensuremath{\text{\,#1}}}       

\newcommand{\tev}{\aunit{Te\kern -0.1em V}\xspace}
\newcommand{\gev}{\aunit{Ge\kern -0.1em V}\xspace}
\newcommand{\mev}{\aunit{Me\kern -0.1em V}\xspace}
\newcommand{\kev}{\aunit{ke\kern -0.1em V}\xspace}
\newcommand{\ev}{\aunit{e\kern -0.1em V}\xspace}
 
\newcommand{\mevc}{\ensuremath{\aunit{Me\kern -0.1em V\!/}c}\xspace}
\newcommand{\gevc}{\ensuremath{\aunit{Ge\kern -0.1em V\!/}c}\xspace}
\newcommand{\mevcc}{\ensuremath{\aunit{Me\kern -0.1em V\!/}c^2}\xspace}
\newcommand{\gevcc}{\ensuremath{\aunit{Ge\kern -0.1em V\!/}c^2}\xspace}

\def\fb   {\ensuremath{\aunit{fb}}\xspace}
\def\invfb   {\ensuremath{\fb^{-1}}\xspace}

\newcommand{\chisq}{\ensuremath{\chi^2}\xspace}
\newcommand{\chisqndf}{\ensuremath{\chi^2/\mathrm{ndf}}\xspace}

\def\gsim{{~\raise.15em\hbox{$>$}\kern-.85em
          \lower.35em\hbox{$\sim$}~}\xspace}
\def\lsim{{~\raise.15em\hbox{$<$}\kern-.85em
          \lower.35em\hbox{$\sim$}~}\xspace}

\def\pt         {\ensuremath{p_{\mathrm{T}}}\xspace}

\def\evtgen     {\mbox{\textsc{EvtGen}}\xspace}

\def\geant      {\mbox{\textsc{Geant4}}\xspace}

\def\photos     {\mbox{\textsc{Photos}}\xspace}
\def\powheg     {\mbox{\textsc{Powheg}}\xspace}
\def\pythia     {\mbox{\textsc{Pythia}}\xspace}

\def\tell1  {TELL1\xspace}
\def\ukl1   {UKL1\xspace}

\newcommand{\lhcborcid}[1]{\href{https://orcid.org/#1}{\hspace*{0.1em}\raisebox{-0.45ex}{\includegraphics[width=1em]{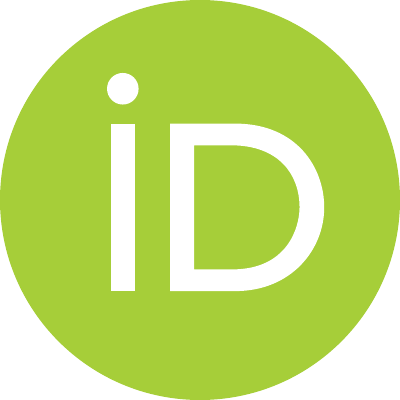}}}}

\hypersetup{
  colorlinks   = true, 
  urlcolor     = blue, 
  linkcolor    = blue, 
  citecolor    = red   
}

\ifEnableSectionTOCLinks
    \usepackage[explicit]{titlesec} 
    
    \let\oldcontentsline\contentsline
    \renewcommand

    \titleformat{\section}{\normalfont\Large\bf}{\hyperlink{tocsection.\thesection}{{\thesection} \parbox[t]{\dimexpr\textwidth-1pc}{#1}}}{1pc}{}

    \titleformat{\subsection}{\normalfont\bf}{\hyperlink{tocsubsection.\thesubsection}{{\thesubsection} \parbox[t]{\dimexpr\textwidth-1pc}{#1}}}{1pc}{}

    \titleformat{name=\section,numberless}[display]{}{}{0pt}{\normalfont\Huge\bfseries #1}
\fi

\usepackage{cite} 
\usepackage{mciteplus}

\usepackage{longtable} 
\begin{document}

\newcommand{\DefaultFigureWidth}{0.7\textwidth}
\newcommand{\mz}{\ensuremath{m_Z}\xspace}
\newcommand{\powhegbox}{{\sc{Powheg-Box}}\xspace}
\newcommand{\powhegewnlo}{{\sc{Powheg}}-ewnlo\xspace}
\newcommand{\powhegewlo}{{\sc{Powheg}}-ewlo\xspace}
\newcommand{\dyturbo}{{\sc{DYTurbo}}\xspace}

\newcommand{\fitndof}{37}
\newcommand{\fitchisq}{43.4}
\newcommand{\mZvalue}{91185.7}
\newcommand{\mZstaticuncertainty}{8.3}
\newcommand{\curvaturevalue}{1.2}
\newcommand{\curvatureuncertainty}{0.1}
\newcommand{\Zfitcorrelation}{0.015}

\newcommand{\totalmZsystematics}{3.9}
\newcommand{\energyoffsetsystematic}{2.6}
\newcommand{\smearingsystematic}{1.8}
\newcommand{\upsilonmasssystematic}{1.5}
\newcommand{\psuedomasssystematic}{0.7}
\newcommand{\upsilonfsrsystematic}{0.6}
\newcommand{\therestsystematic}{1}

\newcommand{\totalmomentumsystematic}{3.6}
\newcommand{\qedfsrsystematic}{1.0}
\newcommand{\pdfsystematic}{0.7}
\newcommand{\efficienciessystematic}{0.1}
\newcommand{\theorytemplatesystematic}{1.0}

\newcommand{\totalmomentumsystematicrelative}{\ensuremath{4\times 10^{-5}}\xspace}

\newcommand{\numZcandidates}{173\,592}
\newcommand{\numUpsilonCandidates}{193\,422}
\newcommand{\numUpsilonLowCandidates}{1\,778\,030}
\newcommand{\numJpsiCandidates}{48\,273}
\newcommand{\approxNumZcandidates}{$174 \times 10^3$}
\newcommand{\approxNumUpsilonCandidates}{$193 \times 10^3$}
\newcommand{\approxNumUpsilonLowCandidates}{$1.8 \times 10^6$}
\newcommand{\approxNumJpsiCandidates}{$48 \times 10^3$}

\newcommand{\NumRunIntervals}{20}

\newcommand{\DirectionalNumMeasurements}{116}
\newcommand{\DirectionalNumMasked}{20}
\newcommand{\DirectionalNDOF}{100}
\newcommand{\DirectionalChisqData}{310}
\newcommand{\DirectionalChisqSim}{158}

\newcommand{\momSmearingFitChiSq}{76}
\newcommand{\momSmearingFitNDOF}{95}
\newcommand{\momSmearingAlpha}{$\alpha=(-0.65\pm 0.16) \times 10^{-4}$}
\newcommand{\momSmearingSigmaOne}{$\sigma_1=(1.98\pm0.07) \times 10^{-3}$}
\newcommand{\momSmearingSigmaTwo}{$\sigma_2=0.147 \pm 0.009\tev^{-1}$}

\newcommand{\SimulationZMass}{\ensuremath{91187.6\mev}\xspace}
\newcommand{\SimulationUpsilonMass}{\ensuremath{9460.3\mev}\xspace}
\newcommand{\SimulationJpsiMass}{\ensuremath{3096.9\mev}\xspace}

\newcommand{\DatasetIntegratedLumi}{1.7}

\renewcommand{\thefootnote}{\fnsymbol{footnote}}
\setcounter{footnote}{1}


\begin{titlepage}
\pagenumbering{roman}

\vspace*{-1.5cm}
\centerline{\large EUROPEAN ORGANIZATION FOR NUCLEAR RESEARCH (CERN)}
\vspace*{1.5cm}
\noindent
\begin{tabular*}{\linewidth}{lc@{\extracolsep{\fill}}r@{\extracolsep{0pt}}}
\ifthenelse{\boolean{pdflatex}}
{\vspace*{-1.5cm}\mbox{\!\!\!\includegraphics[width=.14\textwidth]{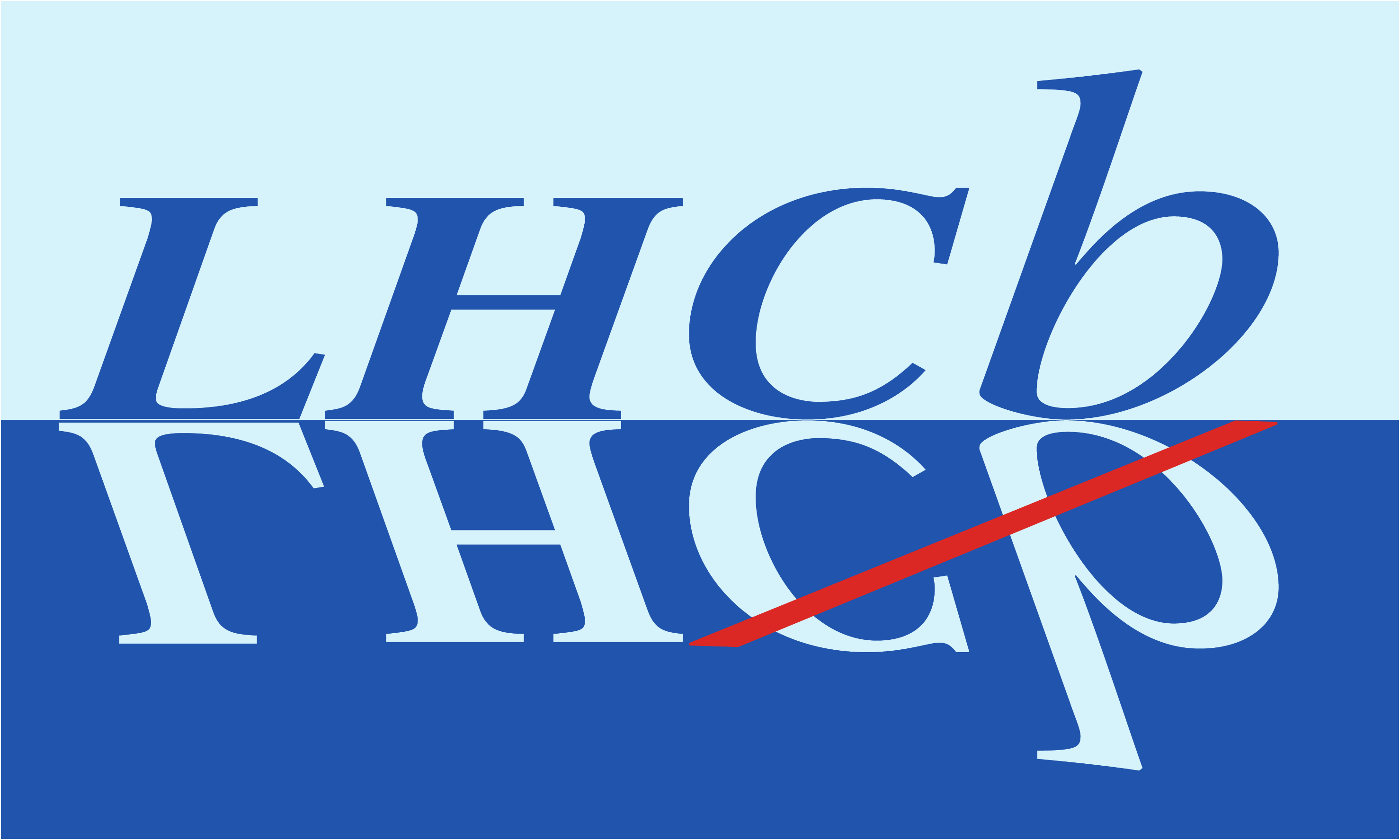}} & &}%
\\
 & & CERN-EP-2025-111 \\  
 & & LHCb-PAPER-2025-008 \\  
 & & \today \\ 
 & & \\
\end{tabular*}

\vspace*{4.0cm}

{\normalfont\bfseries\boldmath\huge
\begin{center}
  \papertitle 
\end{center}
}

\vspace*{2.0cm}

\begin{center}
\paperauthors\footnote{Authors are listed at the end of this paper.}
\end{center}

\vspace{\fill}


\begin{abstract}
\noindent 
The first dedicated \PZ-boson mass measurement at the LHC with $\PZ \to \mu^+\mu^-$ decays is reported. 
The dataset uses proton-proton collisions at a centre-of-mass energy of $13$\,TeV, recorded in 2016 by the LHCb experiment, and corresponds to an integrated luminosity of $1.7$\,fb$^{-1}$.
A template fit to the $\mu^+\mu^-$ mass distribution yields the following result for the $\PZ$-boson mass,
\begin{equation*}
\mz = \mZvalue \pm \mZstaticuncertainty \pm \totalmZsystematics  \mev,
\end{equation*}
where the first uncertainty is statistical and the second systematic. 
This result is consistent with previous measurements and predictions from global electroweak fits. 
\end{abstract}

\vspace*{2.0cm}

\begin{center}
  Published in Phys. Rev. Lett. 135 (2025) 161802
\end{center}

\vspace{\fill}

{\footnotesize 
\centerline{\copyright~\papercopyright. \href{\paperlicenceurl}{\paperlicence}.}}
\vspace*{2mm}

\end{titlepage}


\newpage
\setcounter{page}{2}
\mbox{~}
%
%
%
%


\renewcommand{\thefootnote}{\arabic{footnote}}
\setcounter{footnote}{0}


\cleardoublepage


\pagestyle{plain} 
\setcounter{page}{1}
\pagenumbering{arabic}


\noindent In the Standard Model (SM) of particle physics the electroweak interactions are governed by \mbox{$\text{SU}(2) \times \text{U}(1)$} local gauge symmetry, which is spontaneously broken when the Higgs field acquires a vacuum expectation value $v \approx 246\gev$.
The \PZ boson is a mixture of the $\text{SU}(2)$ and $\text{U}(1)$ gauge fields.
At lowest order, its mass is given by \mbox{$\mz = \frac{1}{2}v(g^2 + g^{\prime 2})^{\frac{1}{2}}$},
where $g$ and $g^{\prime}$ are the $\text{SU}(2)$ and $\text{U}(1)$ gauge couplings, respectively.
Including higher-order corrections and defining \mz in a  relativistic Breit--Wigner lineshape with a mass-dependent width, a SM prediction of \mbox{$\mz =  91204.7 \pm 8.8\mev$} is obtained from a fit to precision electroweak data~\cite{PhysRevD.106.033003}.\footnote{Natural units, where $c=1$, are used throughout this Letter.}

The most precise \mz measurement, combining the data from the four LEP $e^+e^-$ collider experiments, obtained \mbox{$\mz = 91187.6 \pm 2.1\mev$}~\cite{PrecisionZPoleAverages}.
The CDF collaboration reported a measurement of \mz with a precision of $7\mev$, as a validation of their \PW-boson mass measurement~\cite{CDF:2022}. 
A similar study by the CMS collaboration~\cite{CMS:2024lrd}, while presented as not being fully independent of the previously known \mz value, suggests encouraging prospects for a dedicated \mz measurement. 

This Letter reports the first dedicated measurement of \mz at the LHC, using \decay{\PZ}{\mumu}  decays recorded with the \lhcb detector, in proton-proton collisions at a centre-of-mass energy of 13\tev.
The \lhcb detector~\cite{LHCb-DP-2008-001,LHCb-DP-2014-002} instruments the pseudorapidity range \mbox{$2<\eta <5$} with a high-precision tracking system consisting of a silicon-strip vertex detector surrounding the proton-proton interaction region~\cite{LHCb-DP-2014-001}, a large-area silicon-strip detector (the TT) 
located upstream of a dipole magnet with a bending power of about $4{\mathrm{\,T\,m}}$, and three stations of silicon-strip detectors and straw
drift tubes~\cite{LHCb-DP-2017-001} placed downstream of the magnet. 
The tracks are fitted with a Kalman filter that takes into account multiple scattering and corrects for energy loss due to ionisation. 
For particles of charge $q$ and momentum $p$, the typical $q/p$ resolution is $\mathcal{O}(10^{-2})\tev^{-1}$.
This translates to a relative resolution of $\mathcal{O}(1\%)$ on the momenta, typically around $500\gev$, of muons from \PZ-boson decays in LHCb.
Muons are identified by a dedicated system composed of alternating layers of iron and multiwire proportional chambers~\cite{LHCb-DP-2012-002}.
The energies of photons and hadrons are measured with a system of calorimeters. 

The data were recorded in 2016 and correspond to an integrated luminosity of \mbox{$1.7\invfb$}, divided between each of the two magnet-polarity configurations in roughly equal measure.
The online event selection was performed by a trigger system~\cite{LHCb-DP-2012-004,LHCb-DP-2019-001}, and this analysis uses events recorded by single-muon triggers. 
In order to prevent observer bias, the analysis has been carried out by introducing an unknown offset
in the \mz value until the analysis methodology was finalised.

Simulation samples are generated using \pythia~\cite{Sjostrand:2007gs,*Sjostrand:2006za} with a specific \lhcb configuration~\cite{LHCb-PROC-2010-056}.
Decays of heavy particles, such as weak bosons and top quarks, are modelled directly with \pythia, while decays of lighter particles are described by \evtgen~\cite{Lange:2001uf}, in which final-state photon radiation is generated using \photos~\cite{davidson2015photos}.
The interaction of the generated particles with the detector is implemented~\cite{LHCb-PROC-2011-006} using the \geant toolkit~\cite{Allison:2006ve, *Agostinelli:2002hh}.
  
Candidate \decay{\PZ}{\mumu}  decays are formed from pairs of oppositely charged muons, with invariant masses between $86$ and $96$\gev.
The muons must have transverse momenta $\pt > 20$\gev, $p <2$\tev, $2.2 < \eta < 4.4$ and their momenta must have relative uncertainties of less than $6\%$.
Both muons must be consistent with originating from a primary collision vertex~(PV) and have an ``isolation" of less than $40\gev$. The isolation sums the transverse momenta of all other particles for which the quadrature sum of the separation, with respect to the muon, in pseudorapidity and azimuthal angle is below $0.4$.
Around \approxNumZcandidates\ candidates are selected by these requirements, which suppress the hadronic backgrounds to a negligible level with minimal signal inefficiency.

A sample of around \approxNumUpsilonCandidates\ candidate \decay{\Upsilonres}{\mumu} decays is used for the momentum calibration, while roughly \approxNumJpsiCandidates\ candidate \decay{\jpsi}{\mumu} decays are used for an independent cross-check. 
These quarkonium candidates are selected from pairs of oppositely charged muons with $\pt > 5\gev$ and $2.2 < \eta < 4.4$.
The \pt requirements, which are tighter than those typically used in analyses of \Upsilonres and \jpsi decays in LHCb, select a subset of muons that are more similar to those from \PZ-boson decays.
Muon-isolation requirements are not imposed since quarkonium states tend to be produced in hadronic jets~\cite{LHCb-PAPER-2016-064}, but all other muon-quality requirements from the \PZ-boson selection are applied.
The \jpsi candidates are required to have a significant displacement from the PV and therefore originate predominantly from decays of beauty hadrons.

To highlight the importance of the momentum calibration in this analysis, briefly assume the limit of massless muons in which the dimuon mass $m$ is given by $m^2 \simeq 2p^{+}p^{-}(1 - \cos\theta)$ with $p^{\pm}$ denoting the $\mu^{\pm}$ momenta and $\theta$ being the angle between the muons, which is measured with very high precision.
The main challenge is to control systematic biases on the $p^{\pm}$ measurements. 
To a good approximation, these transform the measured momenta of particles with charge $q$ as 
\begin{equation}
\label{eq:mom-form}
    p \rightarrow \left(1 + \alpha + \beta/p -q \delta p \right) p,
\end{equation}
where $\alpha$ is a momentum-scale offset caused by a miscalibration of the bending power of the magnet and/or the tracker length scale; $\beta$ is due to a possible inaccuracy in the ionisation-energy-loss corrections; and $\delta$ is a $q/p$ bias due to a misalignment of the tracking detectors.
These parameters may vary between different detector regions.

A general-purpose momentum calibration for LHCb~\cite{LHCb-DP-2023-003} achieves a relative uncertainty of $3\times 10^{-4}$ for application to a wide range of decays with different lifetimes and final-state particle momenta.
The present analysis requires a dedicated calibration with a higher precision specifically for high-\pt muons and with an emphasis on differences between data and simulation.
This calibration is performed in four stages, the first of which uses the \decay{\PZ}{\mumu}  sample, without an assumption on the \mz value, while the remaining three use the \decay{\Upsilonres}{\mumu} sample.
The first two calibrations apply to the data and the last two subsequently apply to the simulation.

The first momentum calibration estimates the charge-dependent curvature biases ($\delta$) with \decay{\PZ}{\mumu} decays and the  pseudomass method~\cite{Barter:2021npd,LHCb-DP-2023-001}.
Denoting the $\mu^\pm$ momentum and transverse momentum as $p^{\pm}$ and $\pt^{\pm}$, respectively, the two pseudomasses, defined as
\begin{equation}
\mathcal{M}^{\pm} \equiv \sqrt{2p^{\pm}\pt^{\pm}\frac{p^{\mp}}{\pt^{\mp}}(1 - \cos\theta)},
\end{equation}
estimate the dimuon mass under the assumption of a relatively small dimuon transverse momentum.
In intervals of $\eta$, $\phi$, and the magnet polarity, a simultaneous fit of the $\mathcal{M}^+$ and $\mathcal{M}^-$ distributions determines the charge asymmetry in the peak positions, 
which is multiplied by a factor of \mbox{$-\frac{1}{2}(\langle 1/p^+\rangle + \langle 1/p^-\rangle)$} to estimate $\delta$, where $\langle 1/p^\pm\rangle$ are the mean values of the inverse of the muon momenta.
A small bias in the $\delta$ estimates~\cite{Barter:2021npd}, which is an order of magnitude smaller than the dominant misalignment effect, is corrected with simulation as in Ref.~\cite{LHCb-PAPER-2024-028}.
Applying the corrected $\delta$ values to the muon momenta in data improves the \PZ mass resolution by $\mathcal{O}(20\%)$.
Crucially, it is verified that variations as large as $\pm 100\mev$ in the \mz value of \SimulationZMass assumed in the simulation have an effect of less than $1\mev$ on the \mz measurement.

The second momentum calibration accounts for a relative momentum-scale drift of $\mathcal{O}(10^{-4})$ during the data taking.
The $\Upsilonres$-meson sample is divided into \NumRunIntervals\ data-taking periods
and the mass distributions are modelled with a fit function composed of two Crystal~Ball functions~\cite{Skwarnicki:1986xj} for the signal and an exponential for the background. 
The two signal functions share a common peak-position parameter, but differ by a relative-width factor, which varies freely in the fit together with their relative normalisation.
The peak position is parameterised as the known $\Upsilonres$ mass~\cite{PDG2024}, multiplied by $(1 + \alpha)$. 
The momenta in data are corrected according to the \NumRunIntervals\ values for $\alpha$ from these fits, which are shown in Fig.~\ref{fig:rundependent-scale-calibrations} in the Appendix.

The third momentum calibration accounts for a possible underestimation in the simulation of the $\alpha$ variations across the detector. 
Corrections are defined according to 16 discrete ``directions", which correspond to the product of four intervals in the range $2.2 < \eta < 4.4$ and four $\phi$ intervals.
The \Upsilonres-meson sample is divided into \DirectionalNumMeasurements\ subsamples according to the \mup and \mun directions, after folding across the diagonal and, subsequently, removing \DirectionalNumMasked\ subsamples with fewer than 100 candidates in data.
Mass measurements are obtained for these \DirectionalNumMeasurements\ subsamples using the same fit model as in the drift calibration.
The 16 values of $\alpha$ are determined from a \chisq fit in which the measured mass for the subsample with the $\mup (\mun)$ being in the directional interval $i (j)$ is predicted to differ from the known $\Upsilonres$ mass~\cite{PDG2024} by a factor $\sqrt{(1+\alpha_i)(1+\alpha_j)}$.

Figure~\ref{fig:directional-scale-calibrations} in the Appendix shows that in the fit to data, localised $\alpha$ values as large as $\pm 10^{-3}$ are estimated, and the \chisq of the fit is roughly 700 units smaller than that where a single $\alpha$ value is used. 
The variations in the simulation are an order of magnitude smaller, and these are subsequently corrected to match those seen in data.

\begin{figure}[tb]
    \centering
    \includegraphics[width=\DefaultFigureWidth]{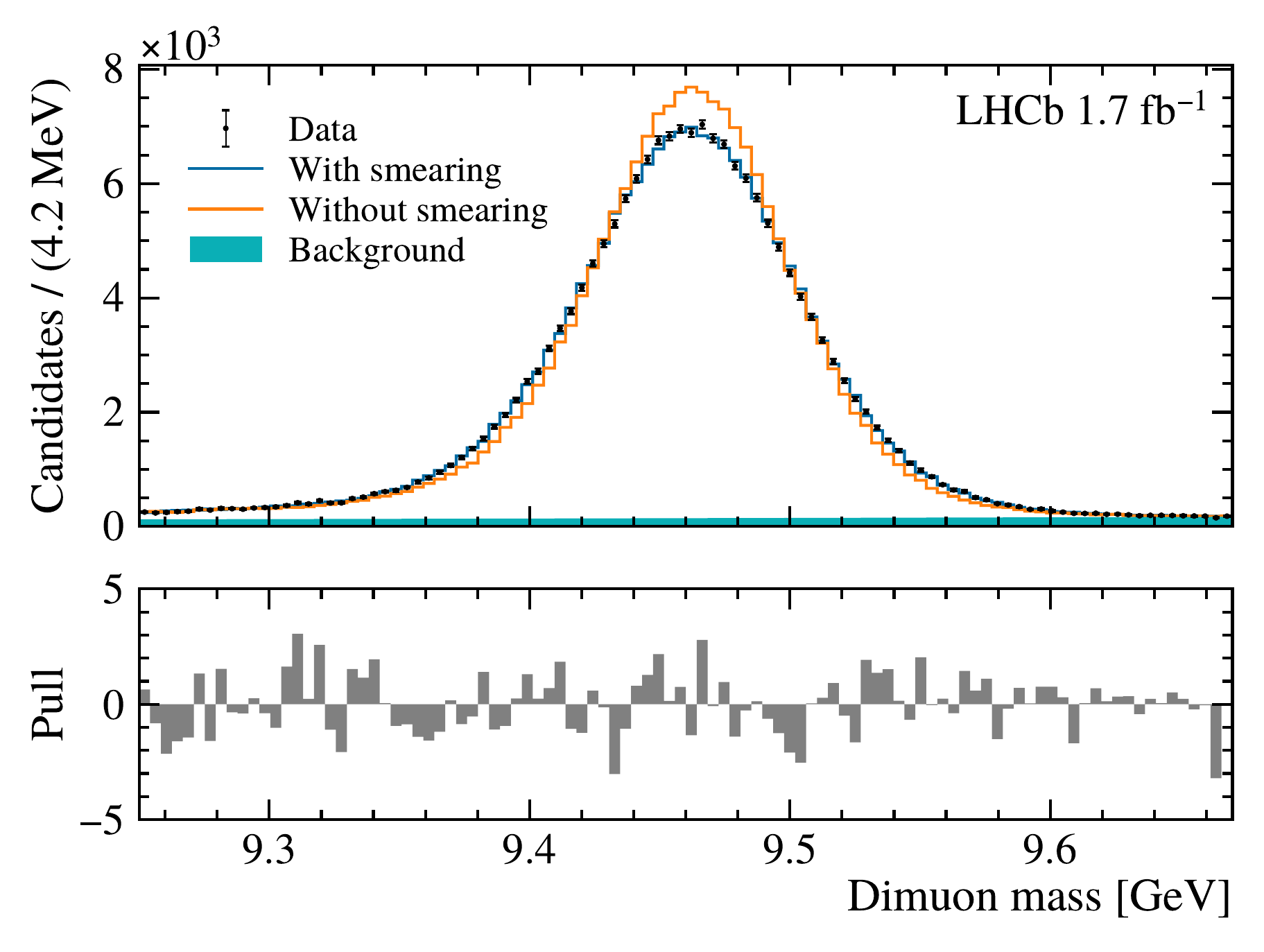}
    \caption{Mass distribution of the \decay{\Upsilonres}{\mumu}candidates, compared to the result of the momentum smearing fit, which includes a background component. The sum of the background and \Upsilonres signal simulation is also shown before momentum smearing.}
    \label{fig:smearing}
\end{figure}

The final, and most important stage of the momentum calibration simultaneously determines an overall $\alpha$ correction and the coefficients $\sigma_{1,2}$ of two momentum smearing factors for the simulation.
The first smearing factor, which accounts for a small underestimation of the multiple scattering in simulation, has the form $(1 + f(\eta)\mathcal{R}_1\sigma_1)$, where $\mathcal{R}_1$ is a random number sampled from a standard normal distribution, and $f(\eta)$ is defined as 1.0 (1.5) for muons with $\eta$ below (above) $3.3$, roughly following an increase in the number of radiation lengths traversed by charged particles.
The second smearing factor, which accounts for an overly idealised $q/p$ resolution in the simulation, has the form $(1 + \mathcal{R}_2 p \sigma_2/\cosh\eta)$, where $\mathcal{R}_2$ is another normally distributed random number.
A fit minimises the \chisq between the $\Upsilonres$ mass distribution in data and simulation, depending on the values of $\alpha$ and $\sigma_{1,2}$
used to smear the momenta in the simulation. 
In this fit, the \Upsilonres mass in the simulation is shifted by $+0.1\mev$ from the default value of \SimulationUpsilonMass to the current known value~\cite{PDG2024}.
The background is modelled with an exponential function with a freely varying slope and normalisation. 
The \chisq of the fit is \momSmearingFitChiSq\ for \momSmearingFitNDOF\ degrees of freedom.
Figure~\ref{fig:smearing} shows the \Upsilonres mass distribution in data, compared to the simulation before and after smearing with the best-fit values of \mbox{\momSmearingAlpha}, \mbox{\momSmearingSigmaOne}~and \mbox{\momSmearingSigmaTwo}. 
Despite the preceding calibrations, the negative $\alpha$ value can be understood as follows: a shift of roughly $-10^{-5}$ is attributed to the change in the \Upsilonres mass, while the remaining shift is attributed to a subtle anticorrelation of around $-12\%$ between $\alpha$ and $\sigma_2$, which is fixed to zero in the earlier stages.

Weights are assigned to the simulated candidates to correct for small inaccuracies in the modelling of the detection efficiency. 
Following the approaches of Refs.~\cite{LHCb-PAPER-2021-024, LHCb-PAPER-2024-028}, these weights are derived from estimates of the  track reconstruction, muon identification and trigger efficiencies.
These corrections are determined from samples of \decay{\PZ}{\mumu}  and \decay{\Upsilonres}{\mumu} decays in $\eta$ and $\phi$ intervals, with parametrisations of the \pt dependence.
The tracking and identification efficiency corrections are typically within 1\% of unity, but the trigger efficiencies are overestimated by up to $\mathcal{O}(10\%)$ at the largest $\eta$ values considered in this analysis. 

\begin{figure}[h]
    \centering
    \includegraphics[width=\DefaultFigureWidth]{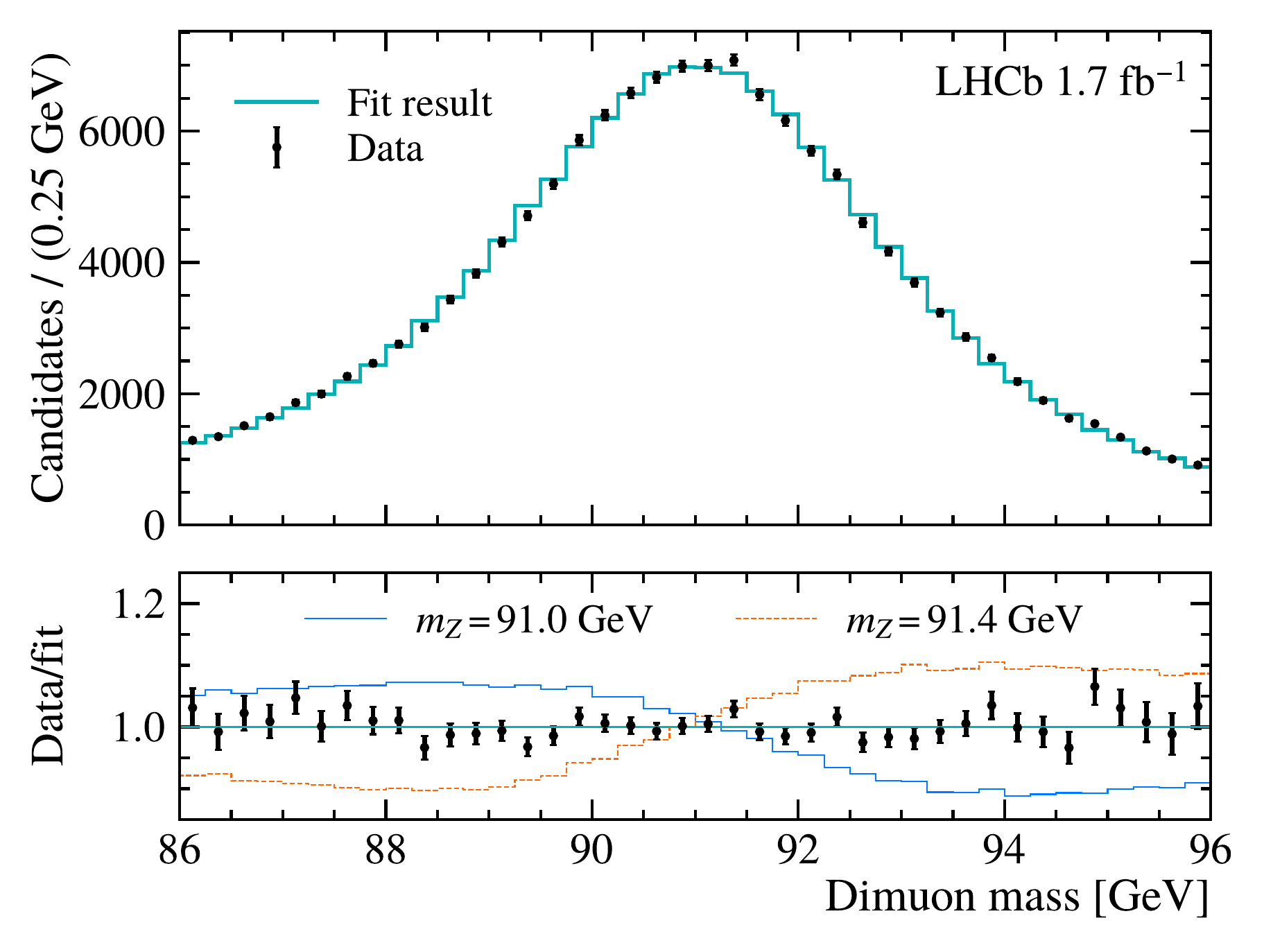}
    \caption{Mass distribution of the \decay{\PZ}{\mumu}  candidates, compared to predictions for two different \mz values and to the results of the fit. The predictions and fit results include the small background component.}
    \label{fig:zfit}
\end{figure}

The \mz measurement is obtained from a \chisq fit to the dimuon mass distribution in 40 intervals between 86 and 96\gev, shown in Fig.~\ref{fig:zfit}. 
Backgrounds from $\decay{\PZ}{\tautau}$, top-quark decays, vector-boson pairs, and hadrons are included, with their total fraction being $\mathcal{O}(10^{-3})$.
The signal and background components are modelled with templates from the simulation samples with the calibrations already described.
The signal normalisation is allowed to vary freely in the fit.

Auxiliary signal samples without detector simulation are produced using a version of \powhegbox~\cite{powheg1,powheg2,powheg3} that includes next-to-leading-order corrections in the strong and electroweak couplings~\cite{POWHEGewZ,POWHEGewZ2}.
The NNPDF3.1~\cite{Ball:2017nwa} parton distribution functions are used, and the samples are processed with \photos~\cite{Barberio:1990ms} for modelling of additional photon radiation and with \pythia~\cite{Sjostrand:2007gs} for simulating the rest of the event. 
Normalised mass distributions are produced for \mz values of 91.1, 91.2 and 91.3\gev in the $G_\mu$ input scheme~\cite{ewschemes}. These are divided by the distribution corresponding to the main simulation sample to make three ratio histograms, each of which is translated to event weights for the main simulation sample, evaluated as a function of the true mass.
The momentum-smearing fit precisely determines $\alpha$ but the two $\sigma$ parameters are less precisely determined from the $\Upsilonres$-meson samples. 
Therefore, $\sigma_2$, which is far more influential than $\sigma_1$ for the higher-momentum muons from \PZ-boson decays, varies freely by a factor $f$ in the fit.
Cubic-spline interpolation is used to continuously vary \mz and $f$ in the fit.
The fit has a minimum \chisqndf of $\fitchisq/\fitndof$ and the best fit values are \mbox{$\mz = \mZvalue \pm \mZstaticuncertainty\mev$} and \mbox{$f = \curvaturevalue \pm \curvatureuncertainty$}, with a correlation coefficient of $\Zfitcorrelation$, where the uncertainties are statistical only. 
Figure~\ref{fig:zfit} shows the result of this fit compared to the distribution in data. 

Table~\ref{tab:unc} summarises the systematic uncertainties on \mz, in which the total is the combination in quadrature of the individual components.
Most sources of systematic uncertainty are predominantly statistical, and will decrease in future analyses of larger datasets.
The largest source of uncertainty is attributed to the momentum calibration, described in more detail in the next paragraph.
The next source of systematic uncertainty is attributed to the statistical precision of the true dimuon mass distributions used to define the weights for the simulated events.
An uncertainty for the electroweak corrections in the signal model is estimated by repeating the fit with a model in which the photon radiation, beyond the electroweak corrections modelled directly by \powheg, is handled by \pythia instead of \photos. The shift in the \mz value, which has a statistical uncertainty of $0.7 \mev$, is assigned as the systematic uncertainty.
The uncertainty due to the parton distribution functions is assessed from the envelope of fit results also including those based on the CT18~\cite{Hou:2019efy} and MSHT20~\cite{Bailey:2020ooq} sets.
The detection efficiency uncertainty is obtained by propagating the statistical uncertainties in the estimated efficiencies. 

The dominant momentum calibration uncertainty is comprised of five lower-level sources that are combined in quadrature.
An uncertainty of \energyoffsetsystematic\mev is evaluated for the detector material description in the simulation. Variations of $\pm 10\%$ in the number of radiation lengths upstream of the magnet are emulated with values of $\beta$, set to zero by default, of $\pm 2\mev$ applied to the simulation.
While the $\beta$ parameter has vanishing influence on the muons from \PZ-boson decays, it influences the determination of the $\alpha$ parameter in the momentum-smearing fit.
An uncertainty of \smearingsystematic\mev results from propagating the covariance of the three parameters from the momentum-smearing fit. 
The propagation of the uncertainty on the known $\Upsilonres$-meson mass~\cite{PDG2024} results in an uncertainty of \upsilonmasssystematic\mev.
An uncertainty of \psuedomasssystematic\mev is estimated by propagating the statistical uncertainties on the $\delta$ bias corrections.
Finally, the difference between the \pythia and \photos descriptions of final-state photon radiation in the \Upsilonres simulation results in an uncertainty of \upsilonfsrsystematic\mev.

\begin{table}[t] 
\centering
\caption{\label{tab:unc}Systematic uncertainties on the \mz measurement.}
\begin{tabular}{lc}
Source & Uncertainty [MeV] \\
\hline 
Momentum calibration & \totalmomentumsystematic \\ 
Theory template statistics & \theorytemplatesystematic \\
Signal final-state radiation & \qedfsrsystematic\\
Parton distribution functions & \pdfsystematic\\
Detection efficiency & \efficienciessystematic\\
\hline 
Total systematic uncertainty& \totalmZsystematics \\
\end{tabular}
\end{table}

Several cross-checks are performed to validate the measurement and its uncertainty.
The \mz fit is repeated with the samples divided into roughly equal-size pairs of subsamples according to the magnet polarity, the number of clusters in the scintillator-preshower detector, the number of tracks in the PV, and the angle between the decay plane and the $y$-component of the magnetic field. The sample is also divided according to the opening angle between the muons, the sign of the difference between the muon pseudorapidities, the dimuon \pt, and the dimuon rapidity. Finally, the sample is split according to eight variables for each muon: the number of TT clusters, track-fit \chisqndf, momentum and \pt, the relative uncertainty on the measured  momentum, isolation, the consistency of the muons with a PV, and pseudorapidity. Figure~\ref{fig:cross_checks} shows that the  \mz values corresponding to the subsamples differ by less than two standard deviations in all cases, where only the statistical uncertainties are accounted for.
The \mz fit is also repeated in four intervals of the angle $\phi$ of the \mun and these are consistent within two standard deviations.

For the momentum-scale calibrations and the momentum-smearing fit, it is verified that repeating the fit, having applied the calibration, results in momentum-scale parameters consistent with zero. 
Several variations in the numbers of $\eta$ and $\phi$ intervals in the third momentum calibration are tested. The largest variation in \mz is $2\mev$, but this corresponds to an extreme reduction in the number of intervals, which causes an obvious deficiency in the simulation to be uncorrected. Therefore, no systematic uncertainty is assigned to this third calibration.
The \mz fit is tested with simulated events as pseudodata, validating the coverage of the uncertainty and ruling out any bias. This test is repeated with a wide range of \mz values in the pseudodata, ruling out any significant nonlinear dependence. 
Reasonable variations in the number of mass intervals in the \mz fit have a negligible effect on the results.
The \mz fit is tested with smaller ($88$ to $94$\gev) and larger ($84$ to $98$\gev) fit ranges than that of the baseline fit, and all results are consistent within two standard deviations.
Finally, the \mz measurement is repeated with two independent samples used in the momentum calibration and its uncertainty assessment. The first is $\jpsi\to\mu^+\mu^-$ decays with the same muon requirements as the default $\Upsilonres$ sample. The second is $\Upsilonres\to\mu^+\mu^-$ decays where both muons satisfy $\pt > 3\gev$ but one fails the default requirement of $\pt > 5\gev$. Figure~\ref{fig:different-calibration-samples} shows that the systematic uncertainties are larger with these alternative samples and the central \mz values are consistent with the baseline result within 2\mev. 

In summary, the first dedicated measurement of the \PZ-boson mass at the \lhc is presented. 
This is based on a template fit to the mass distribution of \decay{\PZ}{\mumu} decays, in a dataset recorded by the \lhcb experiment in 2016, corresponding to an integrated luminosity of $\DatasetIntegratedLumi \invfb$.
The momentum calibration is primarily based on a sample of \decay{\Upsilonres}{\mumu} decays.
The resulting measurement of the \PZ-boson mass is 
\begin{equation*}
\mz = \mZvalue \pm \mZstaticuncertainty \pm \totalmZsystematics  \mev,
\end{equation*}
where the first uncertainty is statistical and the second systematic. 
This result is consistent with previous measurements and it is consistent with, and matches the precision of, the predictions of the global electroweak fits.
The unique design of the LHCb detector means that the uncertainties are expected to be mostly uncorrelated with potential measurements of the \PZ-boson mass by the other LHC experiments.
Since its uncertainty is predominantly statistical, this result demonstrates encouraging prospects for analyses with larger LHCb datasets, with roughly an order of magnitude more data having already been recorded, and roughly two orders of magnitude projected after the high-luminosity LHC phase. Further work will be required to reduce the detector material uncertainty, but it is now conceivable that LHCb could challenge the precision of the LEP \PZ-boson mass determination in the future.


\emph{Data availability:} Data associated to the plots in this publication are made available on the CERN Document Server in Ref.~\cite{LHCb-PAPER-2025-008-cds}.

\section*{Acknowledgements}
%
%
\noindent We express our gratitude to our colleagues in the CERN
accelerator departments for the excellent performance of the LHC. We
thank the technical and administrative staff at the LHCb
institutes.
We acknowledge support from CERN and from the national agencies:
ARC (Australia);
CAPES, CNPq, FAPERJ and FINEP (Brazil); 
MOST and NSFC (China); 
CNRS/IN2P3 (France); 
BMBF, DFG and MPG (Germany); 
INFN (Italy); 
NWO (Netherlands); 
MNiSW and NCN (Poland); 
MCID/IFA (Romania); 
MICIU and AEI (Spain);
SNSF and SER (Switzerland); 
NASU (Ukraine); 
STFC (United Kingdom); 
DOE NP and NSF (USA).
We acknowledge the computing resources that are provided by ARDC (Australia), 
CBPF (Brazil),
CERN, 
IHEP and LZU (China),
IN2P3 (France), 
KIT and DESY (Germany), 
INFN (Italy), 
SURF (Netherlands),
Polish WLCG (Poland),
IFIN-HH (Romania), 
PIC (Spain), CSCS (Switzerland), 
and GridPP (United Kingdom).
We are indebted to the communities behind the multiple open-source
software packages on which we depend.
Individual groups or members have received support from
Key Research Program of Frontier Sciences of CAS, CAS PIFI, CAS CCEPP, 
Fundamental Research Funds for the Central Universities,  and Sci.\ \& Tech.\ Program of Guangzhou (China);
Minciencias (Colombia);
EPLANET, Marie Sk\l{}odowska-Curie Actions, ERC and NextGenerationEU (European Union);
A*MIDEX, ANR, IPhU and Labex P2IO, and R\'{e}gion Auvergne-Rh\^{o}ne-Alpes (France);
Alexander-von-Humboldt Foundation (Germany);
ICSC (Italy); 
Severo Ochoa and Mar\'ia de Maeztu Units of Excellence, GVA, XuntaGal, GENCAT, InTalent-Inditex and Prog.~Atracci\'on Talento CM (Spain);
SRC (Sweden);
the Leverhulme Trust, the Royal Society and UKRI (United Kingdom).


\clearpage
\section*{Appendix}
\label{sec:Supplemental}

\begin{figure}[b]
    \centering
    \includegraphics[width=\DefaultFigureWidth]{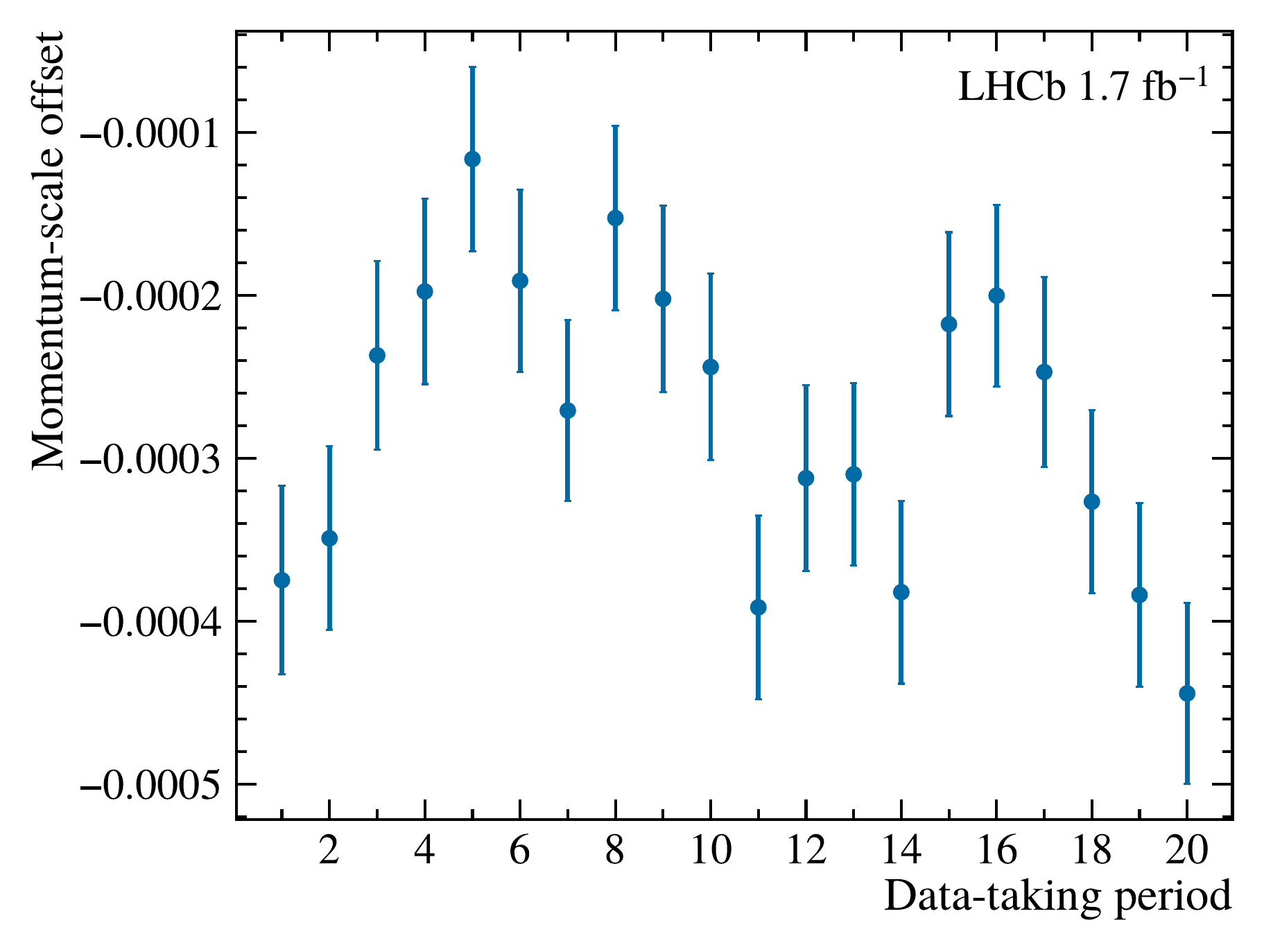}
    \caption{Momentum-scale offset, corresponding to $\alpha$ in Eq.~\ref{eq:mom-form}, in data-taking periods for data.}
    \label{fig:rundependent-scale-calibrations}
\end{figure}

\begin{figure}
    \centering
    \includegraphics[width=\DefaultFigureWidth]{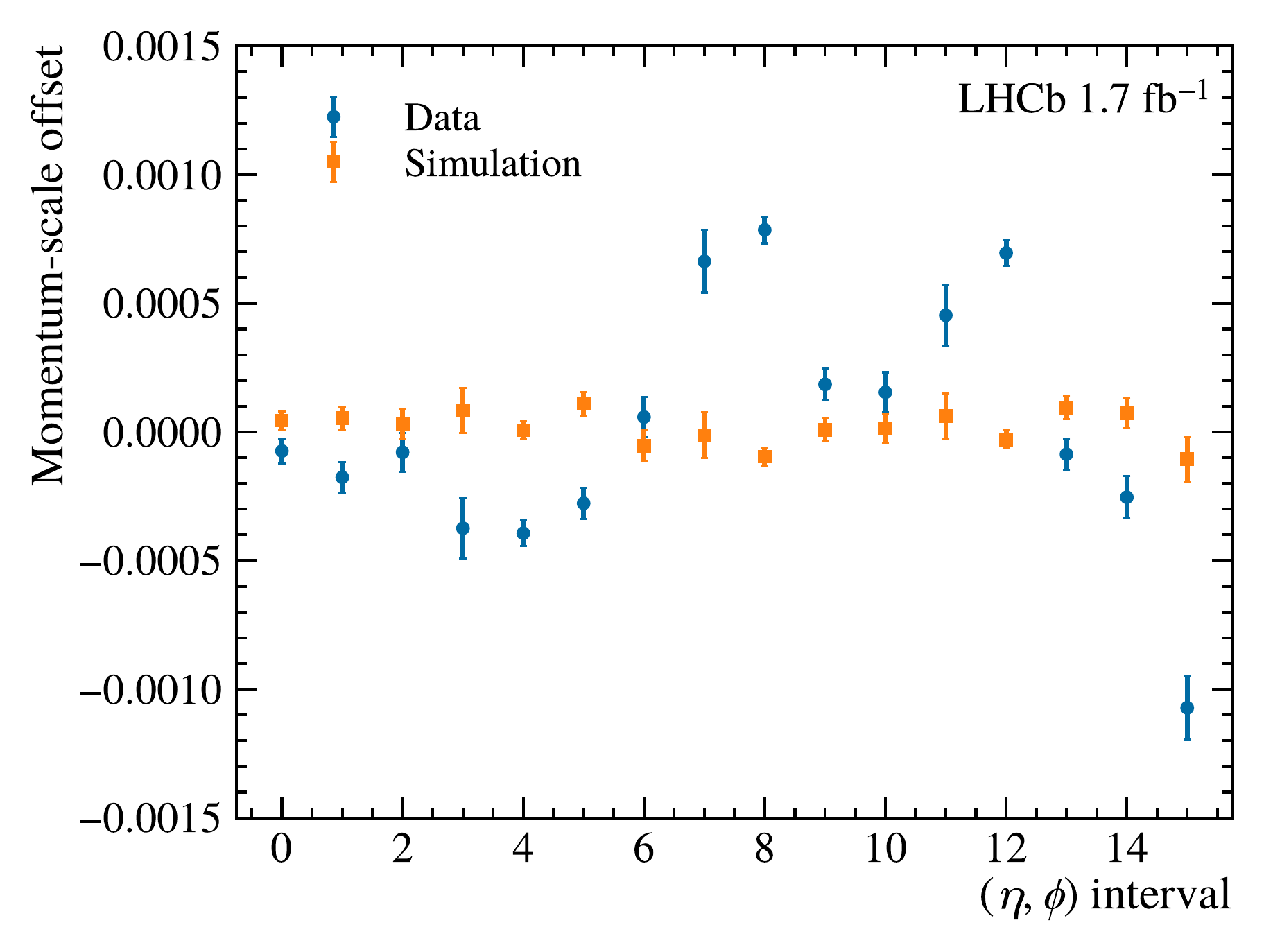}
    \caption{Momentum-scale offsets, corresponding to $\alpha$ in Eq.~\ref{eq:mom-form}, for data and simulation, in intervals of muon pseudorapidity~($\eta$) and azimuthal angle ($\phi$).}
    \label{fig:directional-scale-calibrations}
\end{figure}

\begin{figure}
    \centering
    \includegraphics[width=\DefaultFigureWidth]{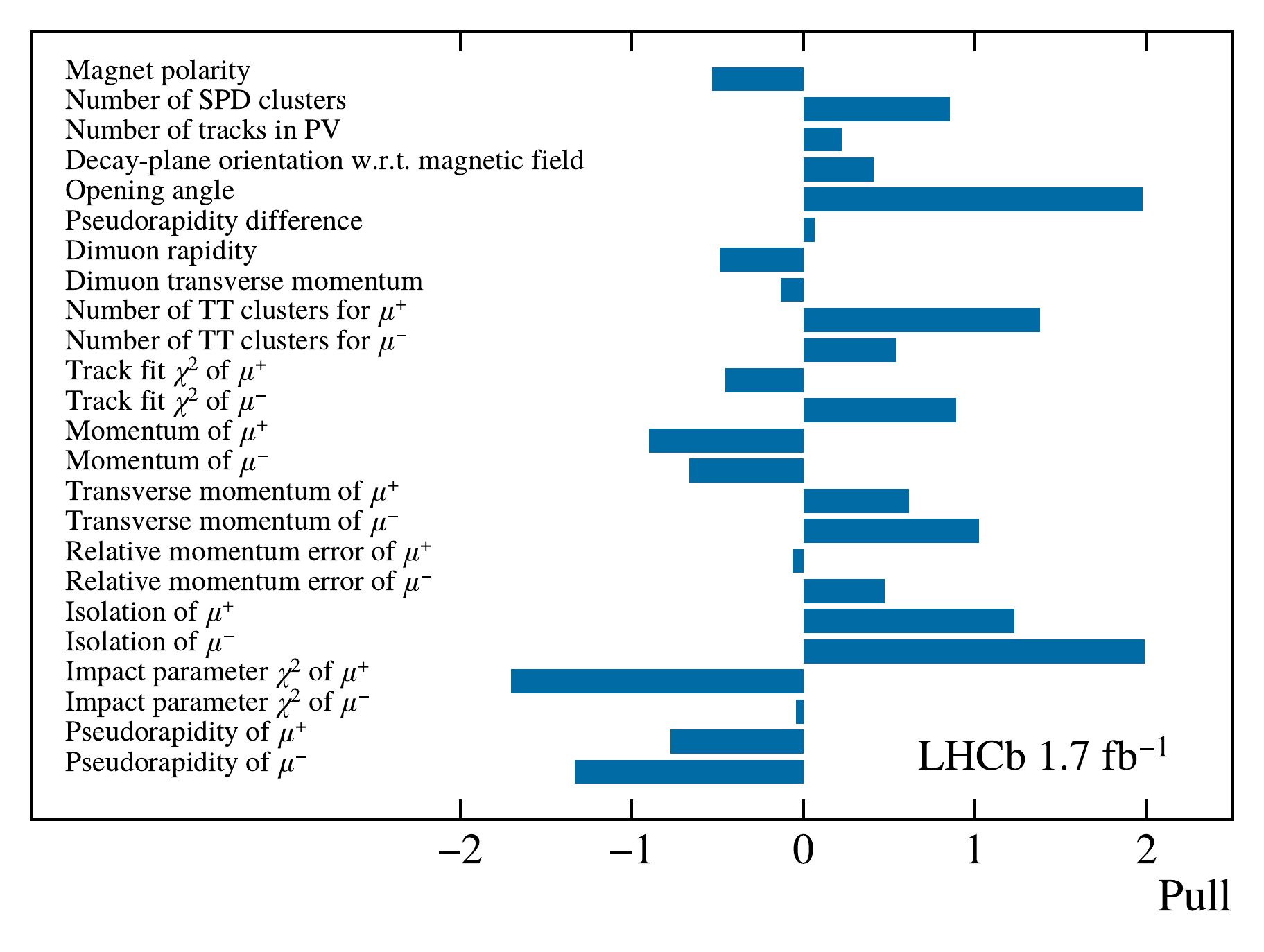}
    \caption{Summary of cross-checks where the data are divided into subsamples of approximately equal size. The pull is defined as the difference between the measurements of a pair of subsamples divided by the sum in quadrature of their statistical uncertainties.}
    \label{fig:cross_checks}
\end{figure}

\begin{figure}
    \centering
      \includegraphics[width=\DefaultFigureWidth]{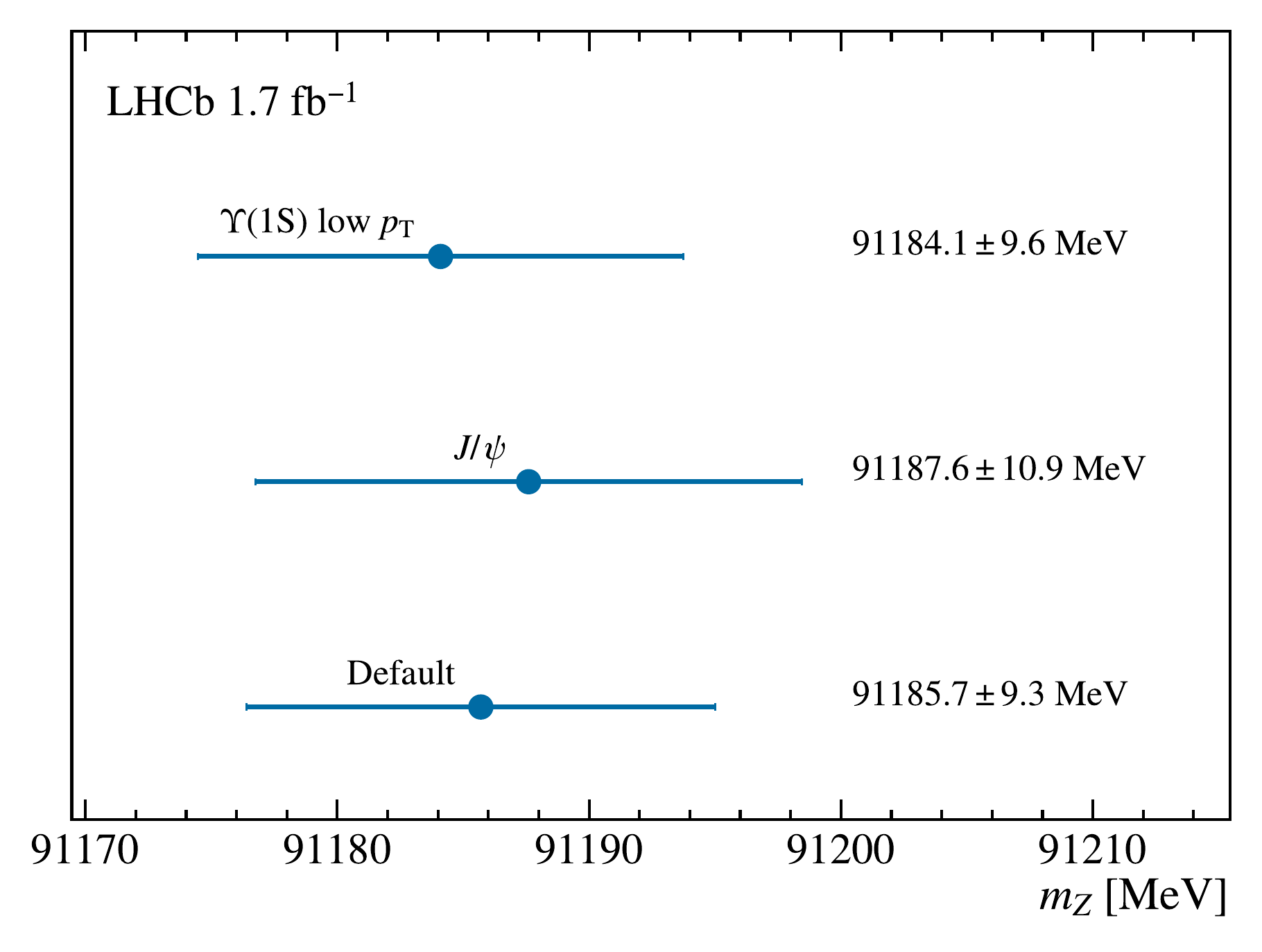}
          \caption{Measurements of \mz with the baseline calibration sample and two alternatives. The uncertainties correspond to the quadrature sum of statistical and systematic sources. For the \jpsi sample, the uncertainty on the known \Upsilonres mass is replaced with that for the \jpsi meson. The final-state-radiation uncertainty is also adapted for the specific \jpsi-meson decay model.}
    \label{fig:different-calibration-samples}
\end{figure}

Figure~\ref{fig:rundependent-scale-calibrations} shows the momentum-scale correction factors in data-taking period intervals. 
Figure~\ref{fig:directional-scale-calibrations} shows the directional momentum-scale offsets for data and simulation.
Figure~\ref{fig:cross_checks} shows the pulls for a set of cross-checks where the samples are divided into two subsamples of roughly equal size. 
Figure~\ref{fig:different-calibration-samples} shows the \mz results based on three different calibration samples.

\clearpage 
\addcontentsline{toc}{section}{References}
\bibliographystyle{LHCb}
\bibliography{main,standard,LHCb-PAPER,LHCb-CONF,LHCb-DP,LHCb-TDR}

\ifx\mcitethebibliography\mciteundefinedmacro
\PackageError{LHCb.bst}{mciteplus.sty has not been loaded}
{This bibstyle requires the use of the mciteplus package.}\fi
\providecommand{\href}[2]{#2}
\begin{mcitethebibliography}{10}
\mciteSetBstSublistMode{n}
\mciteSetBstMaxWidthForm{subitem}{\alph{mcitesubitemcount})}
\mciteSetBstSublistLabelBeginEnd{\mcitemaxwidthsubitemform\space}
{\relax}{\relax}

\bibitem{PhysRevD.106.033003}
J.~de~Blas {\em et~al.}, \ifthenelse{\boolean{articletitles}}{\emph{Global analysis of electroweak data in the standard model}, }{}\href{https://doi.org/10.1103/PhysRevD.106.033003}{Phys.\ Rev.\  \textbf{D106} (2022) 033003}, \href{http://arxiv.org/abs/2112.07274}{{\normalfont\ttfamily arXiv:2112.07274}}\relax
\mciteBstWouldAddEndPuncttrue
\mciteSetBstMidEndSepPunct{\mcitedefaultmidpunct}
{\mcitedefaultendpunct}{\mcitedefaultseppunct}\relax
\EndOfBibitem
\bibitem{PrecisionZPoleAverages}
ALEPH, DELPHI, L3, OPAL, SLD, LEP Electroweak Working Group, SLD Electroweak Group, SLD Heavy Flavour Group, S.~Schael {\em et~al.}, \ifthenelse{\boolean{articletitles}}{\emph{{Precision electroweak measurements on the $Z$ resonance}}, }{}\href{https://doi.org/10.1016/j.physrep.2005.12.006}{Phys.\ Rept.\  \textbf{427} (2006) 257}, \href{http://arxiv.org/abs/hep-ex/0509008}{{\normalfont\ttfamily arXiv:hep-ex/0509008}}\relax
\mciteBstWouldAddEndPuncttrue
\mciteSetBstMidEndSepPunct{\mcitedefaultmidpunct}
{\mcitedefaultendpunct}{\mcitedefaultseppunct}\relax
\EndOfBibitem
\bibitem{CDF:2022}
CDF collaboration, T.~Aaltonen {\em et~al.}, \ifthenelse{\boolean{articletitles}}{\emph{{High-precision measurement of the $W$ boson mass with the CDF II detector}}, }{}\href{https://doi.org/10.1126/science.abk1781}{Science \textbf{376} (2022) 170}\relax
\mciteBstWouldAddEndPuncttrue
\mciteSetBstMidEndSepPunct{\mcitedefaultmidpunct}
{\mcitedefaultendpunct}{\mcitedefaultseppunct}\relax
\EndOfBibitem
\bibitem{CMS:2024lrd}
CMS collaboration, V.~Chekhovsky {\em et~al.}, \ifthenelse{\boolean{articletitles}}{\emph{{High-precision measurement of the W boson mass with the CMS experiment at the LHC}}, }{}\href{http://arxiv.org/abs/2412.13872}{{\normalfont\ttfamily arXiv:2412.13872}}\relax
\mciteBstWouldAddEndPuncttrue
\mciteSetBstMidEndSepPunct{\mcitedefaultmidpunct}
{\mcitedefaultendpunct}{\mcitedefaultseppunct}\relax
\EndOfBibitem
\bibitem{LHCb-DP-2008-001}
LHCb collaboration, A.~A. Alves~Jr.\ {\em et~al.}, \ifthenelse{\boolean{articletitles}}{\emph{{The \lhcb detector at the LHC}}, }{}\href{https://doi.org/10.1088/1748-0221/3/08/S08005}{JINST \textbf{3} (2008) S08005}\relax
\mciteBstWouldAddEndPuncttrue
\mciteSetBstMidEndSepPunct{\mcitedefaultmidpunct}
{\mcitedefaultendpunct}{\mcitedefaultseppunct}\relax
\EndOfBibitem
\bibitem{LHCb-DP-2014-002}
LHCb collaboration, R.~Aaij {\em et~al.}, \ifthenelse{\boolean{articletitles}}{\emph{{LHCb detector performance}}, }{}\href{https://doi.org/10.1142/S0217751X15300227}{Int.\ J.\ Mod.\ Phys.\  \textbf{A30} (2015) 1530022}, \href{http://arxiv.org/abs/1412.6352}{{\normalfont\ttfamily arXiv:1412.6352}}\relax
\mciteBstWouldAddEndPuncttrue
\mciteSetBstMidEndSepPunct{\mcitedefaultmidpunct}
{\mcitedefaultendpunct}{\mcitedefaultseppunct}\relax
\EndOfBibitem
\bibitem{LHCb-DP-2014-001}
R.~Aaij {\em et~al.}, \ifthenelse{\boolean{articletitles}}{\emph{{Performance of the LHCb Vertex Locator}}, }{}\href{https://doi.org/10.1088/1748-0221/9/09/P09007}{JINST \textbf{9} (2014) P09007}, \href{http://arxiv.org/abs/1405.7808}{{\normalfont\ttfamily arXiv:1405.7808}}\relax
\mciteBstWouldAddEndPuncttrue
\mciteSetBstMidEndSepPunct{\mcitedefaultmidpunct}
{\mcitedefaultendpunct}{\mcitedefaultseppunct}\relax
\EndOfBibitem
\bibitem{LHCb-DP-2017-001}
P.~d'Argent {\em et~al.}, \ifthenelse{\boolean{articletitles}}{\emph{{Improved performance of the LHCb Outer Tracker in LHC Run 2}}, }{}\href{https://doi.org/10.1088/1748-0221/12/11/P11016}{JINST \textbf{12} (2017) P11016}, \href{http://arxiv.org/abs/1708.00819}{{\normalfont\ttfamily arXiv:1708.00819}}\relax
\mciteBstWouldAddEndPuncttrue
\mciteSetBstMidEndSepPunct{\mcitedefaultmidpunct}
{\mcitedefaultendpunct}{\mcitedefaultseppunct}\relax
\EndOfBibitem
\bibitem{LHCb-DP-2012-002}
A.~A. Alves~Jr.\ {\em et~al.}, \ifthenelse{\boolean{articletitles}}{\emph{{Performance of the LHCb muon system}}, }{}\href{https://doi.org/10.1088/1748-0221/8/02/P02022}{JINST \textbf{8} (2013) P02022}, \href{http://arxiv.org/abs/1211.1346}{{\normalfont\ttfamily arXiv:1211.1346}}\relax
\mciteBstWouldAddEndPuncttrue
\mciteSetBstMidEndSepPunct{\mcitedefaultmidpunct}
{\mcitedefaultendpunct}{\mcitedefaultseppunct}\relax
\EndOfBibitem
\bibitem{LHCb-DP-2012-004}
R.~Aaij {\em et~al.}, \ifthenelse{\boolean{articletitles}}{\emph{{The \lhcb trigger and its performance in 2011}}, }{}\href{https://doi.org/10.1088/1748-0221/8/04/P04022}{JINST \textbf{8} (2013) P04022}, \href{http://arxiv.org/abs/1211.3055}{{\normalfont\ttfamily arXiv:1211.3055}}\relax
\mciteBstWouldAddEndPuncttrue
\mciteSetBstMidEndSepPunct{\mcitedefaultmidpunct}
{\mcitedefaultendpunct}{\mcitedefaultseppunct}\relax
\EndOfBibitem
\bibitem{LHCb-DP-2019-001}
R.~Aaij {\em et~al.}, \ifthenelse{\boolean{articletitles}}{\emph{{Design and performance of the LHCb trigger and full real-time reconstruction in Run 2 of the LHC}}, }{}\href{https://doi.org/10.1088/1748-0221/14/04/P04013}{JINST \textbf{14} (2019) P04013}, \href{http://arxiv.org/abs/1812.10790}{{\normalfont\ttfamily arXiv:1812.10790}}\relax
\mciteBstWouldAddEndPuncttrue
\mciteSetBstMidEndSepPunct{\mcitedefaultmidpunct}
{\mcitedefaultendpunct}{\mcitedefaultseppunct}\relax
\EndOfBibitem
\bibitem{Sjostrand:2007gs}
T.~Sj\"{o}strand, S.~Mrenna, and P.~Skands, \ifthenelse{\boolean{articletitles}}{\emph{{A brief introduction to PYTHIA 8.1}}, }{}\href{https://doi.org/10.1016/j.cpc.2008.01.036}{Comput.\ Phys.\ Commun.\  \textbf{178} (2008) 852}, \href{http://arxiv.org/abs/0710.3820}{{\normalfont\ttfamily arXiv:0710.3820}}\relax
\mciteBstWouldAddEndPuncttrue
\mciteSetBstMidEndSepPunct{\mcitedefaultmidpunct}
{\mcitedefaultendpunct}{\mcitedefaultseppunct}\relax
\EndOfBibitem
\bibitem{Sjostrand:2006za}
T.~Sj\"{o}strand, S.~Mrenna, and P.~Skands, \ifthenelse{\boolean{articletitles}}{\emph{{PYTHIA 6.4 physics and manual}}, }{}\href{https://doi.org/10.1088/1126-6708/2006/05/026}{JHEP \textbf{05} (2006) 026}, \href{http://arxiv.org/abs/hep-ph/0603175}{{\normalfont\ttfamily arXiv:hep-ph/0603175}}\relax
\mciteBstWouldAddEndPuncttrue
\mciteSetBstMidEndSepPunct{\mcitedefaultmidpunct}
{\mcitedefaultendpunct}{\mcitedefaultseppunct}\relax
\EndOfBibitem
\bibitem{LHCb-PROC-2010-056}
I.~Belyaev {\em et~al.}, \ifthenelse{\boolean{articletitles}}{\emph{{Handling of the generation of primary events in Gauss, the LHCb simulation framework}}, }{}\href{https://doi.org/10.1088/1742-6596/331/3/032047}{J.\ Phys.\ Conf.\ Ser.\  \textbf{331} (2011) 032047}\relax
\mciteBstWouldAddEndPuncttrue
\mciteSetBstMidEndSepPunct{\mcitedefaultmidpunct}
{\mcitedefaultendpunct}{\mcitedefaultseppunct}\relax
\EndOfBibitem
\bibitem{Lange:2001uf}
D.~J. Lange, \ifthenelse{\boolean{articletitles}}{\emph{{The EvtGen particle decay simulation package}}, }{}\href{https://doi.org/10.1016/S0168-9002(01)00089-4}{Nucl.\ Instrum.\ Meth.\  \textbf{A462} (2001) 152}\relax
\mciteBstWouldAddEndPuncttrue
\mciteSetBstMidEndSepPunct{\mcitedefaultmidpunct}
{\mcitedefaultendpunct}{\mcitedefaultseppunct}\relax
\EndOfBibitem
\bibitem{davidson2015photos}
N.~Davidson, T.~Przedzinski, and Z.~Was, \ifthenelse{\boolean{articletitles}}{\emph{{PHOTOS interface in C++: Technical and physics documentation}}, }{}\href{https://doi.org/https://doi.org/10.1016/j.cpc.2015.09.013}{Comp.\ Phys.\ Comm.\  \textbf{199} (2016) 86}, \href{http://arxiv.org/abs/1011.0937}{{\normalfont\ttfamily arXiv:1011.0937}}\relax
\mciteBstWouldAddEndPuncttrue
\mciteSetBstMidEndSepPunct{\mcitedefaultmidpunct}
{\mcitedefaultendpunct}{\mcitedefaultseppunct}\relax
\EndOfBibitem
\bibitem{LHCb-PROC-2011-006}
M.~Clemencic {\em et~al.}, \ifthenelse{\boolean{articletitles}}{\emph{{The \lhcb simulation application, Gauss: Design, evolution and experience}}, }{}\href{https://doi.org/10.1088/1742-6596/331/3/032023}{J.\ Phys.\ Conf.\ Ser.\  \textbf{331} (2011) 032023}\relax
\mciteBstWouldAddEndPuncttrue
\mciteSetBstMidEndSepPunct{\mcitedefaultmidpunct}
{\mcitedefaultendpunct}{\mcitedefaultseppunct}\relax
\EndOfBibitem
\bibitem{Allison:2006ve}
Geant4 collaboration, J.~Allison {\em et~al.}, \ifthenelse{\boolean{articletitles}}{\emph{{Geant4 developments and applications}}, }{}\href{https://doi.org/10.1109/TNS.2006.869826}{IEEE Trans.\ Nucl.\ Sci.\  \textbf{53} (2006) 270}\relax
\mciteBstWouldAddEndPuncttrue
\mciteSetBstMidEndSepPunct{\mcitedefaultmidpunct}
{\mcitedefaultendpunct}{\mcitedefaultseppunct}\relax
\EndOfBibitem
\bibitem{Agostinelli:2002hh}
Geant4 collaboration, S.~Agostinelli {\em et~al.}, \ifthenelse{\boolean{articletitles}}{\emph{{Geant4: A simulation toolkit}}, }{}\href{https://doi.org/10.1016/S0168-9002(03)01368-8}{Nucl.\ Instrum.\ Meth.\  \textbf{A506} (2003) 250}\relax
\mciteBstWouldAddEndPuncttrue
\mciteSetBstMidEndSepPunct{\mcitedefaultmidpunct}
{\mcitedefaultendpunct}{\mcitedefaultseppunct}\relax
\EndOfBibitem
\bibitem{LHCb-PAPER-2016-064}
LHCb collaboration, R.~Aaij {\em et~al.}, \ifthenelse{\boolean{articletitles}}{\emph{{Study of \jpsi production in jets}}, }{}\href{https://doi.org/10.1103/PhysRevLett.118.192001}{Phys.\ Rev.\ Lett.\  \textbf{118} (2017) 192001}, \href{http://arxiv.org/abs/1701.05116}{{\normalfont\ttfamily arXiv:1701.05116}}\relax
\mciteBstWouldAddEndPuncttrue
\mciteSetBstMidEndSepPunct{\mcitedefaultmidpunct}
{\mcitedefaultendpunct}{\mcitedefaultseppunct}\relax
\EndOfBibitem
\bibitem{LHCb-DP-2023-003}
LHCb collaboration, R.~Aaij {\em et~al.}, \ifthenelse{\boolean{articletitles}}{\emph{{Momentum scale calibration of the LHCb spectrometer}}, }{}\href{https://doi.org/10.1088/1748-0221/19/02/P02008}{JINST \textbf{19} (2024) P02008}, \href{http://arxiv.org/abs/2312.01772}{{\normalfont\ttfamily arXiv:2312.01772}}\relax
\mciteBstWouldAddEndPuncttrue
\mciteSetBstMidEndSepPunct{\mcitedefaultmidpunct}
{\mcitedefaultendpunct}{\mcitedefaultseppunct}\relax
\EndOfBibitem
\bibitem{Barter:2021npd}
W.~Barter, M.~Pili, and M.~Vesterinen, \ifthenelse{\boolean{articletitles}}{\emph{{A simple method to determine curvature biases in track reconstruction in hadron collider experiments}}, }{}\href{https://doi.org/10.1140/epjc/s10052-021-09016-9}{Eur.\ Phys.\ J.\  \textbf{C81} (2021) 251}, \href{http://arxiv.org/abs/2101.05675}{{\normalfont\ttfamily arXiv:2101.05675}}\relax
\mciteBstWouldAddEndPuncttrue
\mciteSetBstMidEndSepPunct{\mcitedefaultmidpunct}
{\mcitedefaultendpunct}{\mcitedefaultseppunct}\relax
\EndOfBibitem
\bibitem{LHCb-DP-2023-001}
LHCb collaboration, R.~Aaij {\em et~al.}, \ifthenelse{\boolean{articletitles}}{\emph{{Charge-dependent curvature-bias corrections using a pseudomass method}}, }{}\href{https://doi.org/10.1088/1748-0221/19/03/P03010}{JINST \textbf{19} (2024) P03010}, \href{http://arxiv.org/abs/2311.04670}{{\normalfont\ttfamily arXiv:2311.04670}}\relax
\mciteBstWouldAddEndPuncttrue
\mciteSetBstMidEndSepPunct{\mcitedefaultmidpunct}
{\mcitedefaultendpunct}{\mcitedefaultseppunct}\relax
\EndOfBibitem
\bibitem{LHCb-PAPER-2024-028}
LHCb collaboration, R.~Aaij {\em et~al.}, \ifthenelse{\boolean{articletitles}}{\emph{{Measurement of the effective leptonic weak mixing angle}}, }{}\href{https://doi.org/10.1007/JHEP12(2024)026}{{JHEP} \textbf{{12}} (2024) 026}, \href{http://arxiv.org/abs/2410.02502}{{\normalfont\ttfamily arXiv:2410.02502}}\relax
\mciteBstWouldAddEndPuncttrue
\mciteSetBstMidEndSepPunct{\mcitedefaultmidpunct}
{\mcitedefaultendpunct}{\mcitedefaultseppunct}\relax
\EndOfBibitem
\bibitem{Skwarnicki:1986xj}
T.~Skwarnicki, {\em {A study of the radiative cascade transitions between the Upsilon-prime and Upsilon resonances}}, PhD thesis, Institute of Nuclear Physics, Krakow, 1986, {\href{http://inspirehep.net/record/230779/}{DESY-F31-86-02}}\relax
\mciteBstWouldAddEndPuncttrue
\mciteSetBstMidEndSepPunct{\mcitedefaultmidpunct}
{\mcitedefaultendpunct}{\mcitedefaultseppunct}\relax
\EndOfBibitem
\bibitem{PDG2024}
Particle Data Group, S.~Navas {\em et~al.}, \ifthenelse{\boolean{articletitles}}{\emph{{\href{http://pdg.lbl.gov/}{Review of particle physics}}}, }{}\href{https://doi.org/10.1103/PhysRevD.110.030001}{Phys.\ Rev.\  \textbf{D110} (2024) 030001}\relax
\mciteBstWouldAddEndPuncttrue
\mciteSetBstMidEndSepPunct{\mcitedefaultmidpunct}
{\mcitedefaultendpunct}{\mcitedefaultseppunct}\relax
\EndOfBibitem
\bibitem{LHCb-PAPER-2021-024}
LHCb collaboration, R.~Aaij {\em et~al.}, \ifthenelse{\boolean{articletitles}}{\emph{{Measurement of the $W$ boson mass}}, }{}\href{https://doi.org/10.1007/JHEP01(2022)036}{JHEP \textbf{01} (2022) 036}, \href{http://arxiv.org/abs/2109.01113}{{\normalfont\ttfamily arXiv:2109.01113}}\relax
\mciteBstWouldAddEndPuncttrue
\mciteSetBstMidEndSepPunct{\mcitedefaultmidpunct}
{\mcitedefaultendpunct}{\mcitedefaultseppunct}\relax
\EndOfBibitem
\bibitem{powheg1}
P.~Nason, \ifthenelse{\boolean{articletitles}}{\emph{{A new method for combining NLO QCD with shower Monte Carlo algorithms}}, }{}\href{https://doi.org/10.1088/1126-6708/2004/11/040}{JHEP \textbf{11} (2004) 040}, \href{http://arxiv.org/abs/hep-ph/0409146}{{\normalfont\ttfamily arXiv:hep-ph/0409146}}\relax
\mciteBstWouldAddEndPuncttrue
\mciteSetBstMidEndSepPunct{\mcitedefaultmidpunct}
{\mcitedefaultendpunct}{\mcitedefaultseppunct}\relax
\EndOfBibitem
\bibitem{powheg2}
S.~Frixione, P.~Nason, and C.~Oleari, \ifthenelse{\boolean{articletitles}}{\emph{{Matching NLO QCD computations with parton shower simulations: the POWHEG method}}, }{}\href{https://doi.org/10.1088/1126-6708/2007/11/070}{JHEP \textbf{11} (2007) 070}, \href{http://arxiv.org/abs/0709.2092}{{\normalfont\ttfamily arXiv:0709.2092}}\relax
\mciteBstWouldAddEndPuncttrue
\mciteSetBstMidEndSepPunct{\mcitedefaultmidpunct}
{\mcitedefaultendpunct}{\mcitedefaultseppunct}\relax
\EndOfBibitem
\bibitem{powheg3}
S.~Alioli, P.~Nason, C.~Oleari, and E.~Re, \ifthenelse{\boolean{articletitles}}{\emph{{A general framework for implementing NLO calculations in shower Monte Carlo programs: the POWHEG BOX}}, }{}\href{https://doi.org/10.1007/JHEP06(2010)043}{JHEP \textbf{06} (2010) 043}, \href{http://arxiv.org/abs/1002.2581}{{\normalfont\ttfamily arXiv:1002.2581}}\relax
\mciteBstWouldAddEndPuncttrue
\mciteSetBstMidEndSepPunct{\mcitedefaultmidpunct}
{\mcitedefaultendpunct}{\mcitedefaultseppunct}\relax
\EndOfBibitem
\bibitem{POWHEGewZ}
L.~Barze {\em et~al.}, \ifthenelse{\boolean{articletitles}}{\emph{{Neutral current Drell--Yan with combined QCD and electroweak corrections in the POWHEG BOX}}, }{}\href{https://doi.org/10.1140/epjc/s10052-013-2474-y}{Eur.\ Phys.\ J.\  \textbf{C73} (2013) 2474}, \href{http://arxiv.org/abs/1302.4606}{{\normalfont\ttfamily arXiv:1302.4606}}\relax
\mciteBstWouldAddEndPuncttrue
\mciteSetBstMidEndSepPunct{\mcitedefaultmidpunct}
{\mcitedefaultendpunct}{\mcitedefaultseppunct}\relax
\EndOfBibitem
\bibitem{POWHEGewZ2}
M.~Chiesa, C.~L. Del~Pio, and F.~Piccinini, \ifthenelse{\boolean{articletitles}}{\emph{{On electroweak corrections to neutral current Drell\textendash{}Yan with the POWHEG BOX}}, }{}\href{https://doi.org/10.1140/epjc/s10052-024-12908-1}{Eur.\ Phys.\ J.\  \textbf{C84} (2024) 539}, \href{http://arxiv.org/abs/2402.14659}{{\normalfont\ttfamily arXiv:2402.14659}}\relax
\mciteBstWouldAddEndPuncttrue
\mciteSetBstMidEndSepPunct{\mcitedefaultmidpunct}
{\mcitedefaultendpunct}{\mcitedefaultseppunct}\relax
\EndOfBibitem
\bibitem{Ball:2017nwa}
NNPDF collaboration, R.~D. Ball {\em et~al.}, \ifthenelse{\boolean{articletitles}}{\emph{{Parton distributions from high-precision collider data}}, }{}\href{https://doi.org/10.1140/epjc/s10052-017-5199-5}{Eur.\ Phys.\ J.\  \textbf{C77} (2017) 663}, \href{http://arxiv.org/abs/1706.00428}{{\normalfont\ttfamily arXiv:1706.00428}}\relax
\mciteBstWouldAddEndPuncttrue
\mciteSetBstMidEndSepPunct{\mcitedefaultmidpunct}
{\mcitedefaultendpunct}{\mcitedefaultseppunct}\relax
\EndOfBibitem
\bibitem{Barberio:1990ms}
E.~Barberio, B.~van Eijk, and Z.~Was, \ifthenelse{\boolean{articletitles}}{\emph{{PHOTOS: A universal Monte Carlo for QED radiative corrections in decays}}, }{}\href{https://doi.org/10.1016/0010-4655(91)90012-A}{Comput.\ Phys.\ Commun.\  \textbf{66} (1991) 115}\relax
\mciteBstWouldAddEndPuncttrue
\mciteSetBstMidEndSepPunct{\mcitedefaultmidpunct}
{\mcitedefaultendpunct}{\mcitedefaultseppunct}\relax
\EndOfBibitem
\bibitem{ewschemes}
M.~Chiesa, F.~Piccinini, and A.~Vicini, \ifthenelse{\boolean{articletitles}}{\emph{{Direct determination of $\sin^2 \theta^\ell_{\rm {eff}}$ at hadron colliders}}, }{}\href{https://doi.org/10.1103/PhysRevD.100.071302}{Phys.\ Rev.\  \textbf{D100} (2019) 071302}, \href{http://arxiv.org/abs/1906.11569}{{\normalfont\ttfamily arXiv:1906.11569}}\relax
\mciteBstWouldAddEndPuncttrue
\mciteSetBstMidEndSepPunct{\mcitedefaultmidpunct}
{\mcitedefaultendpunct}{\mcitedefaultseppunct}\relax
\EndOfBibitem
\bibitem{Hou:2019efy}
T.-J. Hou {\em et~al.}, \ifthenelse{\boolean{articletitles}}{\emph{{New CTEQ global analysis of quantum chromodynamics with high-precision data from the LHC}}, }{}\href{https://doi.org/10.1103/PhysRevD.103.014013}{Phys.\ Rev.\  \textbf{D103} (2021) 014013}, \href{http://arxiv.org/abs/1912.10053}{{\normalfont\ttfamily arXiv:1912.10053}}\relax
\mciteBstWouldAddEndPuncttrue
\mciteSetBstMidEndSepPunct{\mcitedefaultmidpunct}
{\mcitedefaultendpunct}{\mcitedefaultseppunct}\relax
\EndOfBibitem
\bibitem{Bailey:2020ooq}
S.~Bailey {\em et~al.}, \ifthenelse{\boolean{articletitles}}{\emph{{Parton distributions from LHC, HERA, Tevatron and fixed target data: MSHT20 PDFs}}, }{}\href{https://doi.org/10.1140/epjc/s10052-021-09057-0}{Eur.\ Phys.\ J.\  \textbf{C81} (2021) 341}, \href{http://arxiv.org/abs/2012.04684}{{\normalfont\ttfamily arXiv:2012.04684}}\relax
\mciteBstWouldAddEndPuncttrue
\mciteSetBstMidEndSepPunct{\mcitedefaultmidpunct}
{\mcitedefaultendpunct}{\mcitedefaultseppunct}\relax
\EndOfBibitem
\bibitem{LHCb-PAPER-2025-008-cds}
\url{https://cds.cern.ch/record/2932727}\relax
\mciteBstWouldAddEndPuncttrue
\mciteSetBstMidEndSepPunct{\mcitedefaultmidpunct}
{\mcitedefaultendpunct}{\mcitedefaultseppunct}\relax
\EndOfBibitem
\end{mcitethebibliography}
 
\clearpage 
\centerline
{\large\bf LHCb collaboration}
\begin
{flushleft}
\small
R.~Aaij$^{38}$\lhcborcid{0000-0003-0533-1952},
A.S.W.~Abdelmotteleb$^{57}$\lhcborcid{0000-0001-7905-0542},
C.~Abellan~Beteta$^{51}$\lhcborcid{0009-0009-0869-6798},
F.~Abudin{\'e}n$^{57}$\lhcborcid{0000-0002-6737-3528},
T.~Ackernley$^{61}$\lhcborcid{0000-0002-5951-3498},
A. A. ~Adefisoye$^{69}$\lhcborcid{0000-0003-2448-1550},
B.~Adeva$^{47}$\lhcborcid{0000-0001-9756-3712},
M.~Adinolfi$^{55}$\lhcborcid{0000-0002-1326-1264},
P.~Adlarson$^{84}$\lhcborcid{0000-0001-6280-3851},
C.~Agapopoulou$^{14}$\lhcborcid{0000-0002-2368-0147}
% ...

{\footnotesize \it

% ...
$^{14}$Universit{\'e} Paris-Saclay, CNRS/IN2P3, IJCLab, Orsay, France\\
% ...
$^{38}$Nikhef National Institute for Subatomic Physics, Amsterdam, Netherlands\\
% ...
$^{47}$Instituto Galego de F{\'\i}sica de Altas Enerx{\'\i}as (IGFAE), Universidade de Santiago de Compostela, Santiago de Compostela, Spain\\
% ...
$^{51}$Physik-Institut, Universit{\"a}t Z{\"u}rich, Z{\"u}rich, Switzerland\\
% ...
$^{55}$H.H. Wills Physics Laboratory, University of Bristol, Bristol, United Kingdom\\
% ...
$^{57}$Department of Physics, University of Warwick, Coventry, United Kingdom\\
% ...
$^{60}$School of Physics and Astronomy, University of Glasgow, Glasgow, United Kingdom\\
$^{61}$Oliver Lodge Laboratory, University of Liverpool, Liverpool, United Kingdom\\
% ...
$^{69}$Syracuse University, Syracuse, NY, United States\\
% ...
$^{84}$Department of Physics and Astronomy, Uppsala University, Uppsala, Sweden, associated to $^{60}$\\
% ...
}
\end{flushleft}
\end{document}